\documentclass[twoside,twocolumn,9pt]{article}
\usepackage{extsizes}
\usepackage[super,sort&compress,comma]{natbib} 
\usepackage[version=3]{mhchem}
\usepackage[left=1.5cm, right=1.5cm, top=1.785cm, bottom=2.0cm]{geometry}
\usepackage{balance}
\usepackage{widetext}
\usepackage{times,mathptmx}
\usepackage{sectsty}
\usepackage{graphicx} 
\usepackage{lastpage}
\usepackage[format=plain,justification=raggedright,singlelinecheck=false,font={stretch=1.125,small,sf},labelfont=bf,labelsep=space]{caption}
\usepackage{float}
\usepackage{fancyhdr}
\usepackage{fnpos}
\usepackage[english]{babel}
\usepackage{array}
\usepackage{droidsans}
\usepackage{charter}
\usepackage[T1]{fontenc}
\usepackage[usenames,dvipsnames]{xcolor}
\usepackage{setspace}
\usepackage[compact]{titlesec}

\usepackage{epstopdf}

\definecolor{cream}{RGB}{222,217,201}

\begin{document}

\pagestyle{fancy}
\thispagestyle{plain}
\fancypagestyle{plain}{

\fancyhead[C]{\includegraphics[width=18.5cm]{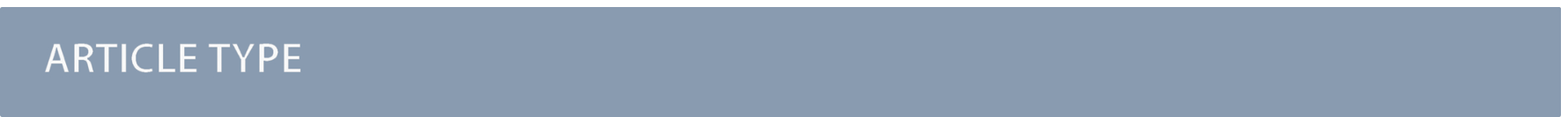}}
\fancyhead[L]{\hspace{0cm}\vspace{1.5cm}\includegraphics[height=30pt]{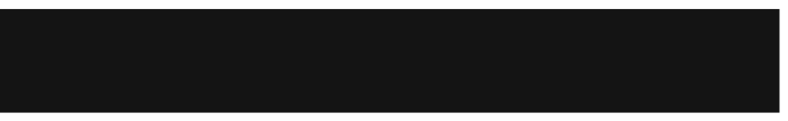}}
\fancyhead[R]{\hspace{0cm}\vspace{1.7cm}\includegraphics[height=55pt]{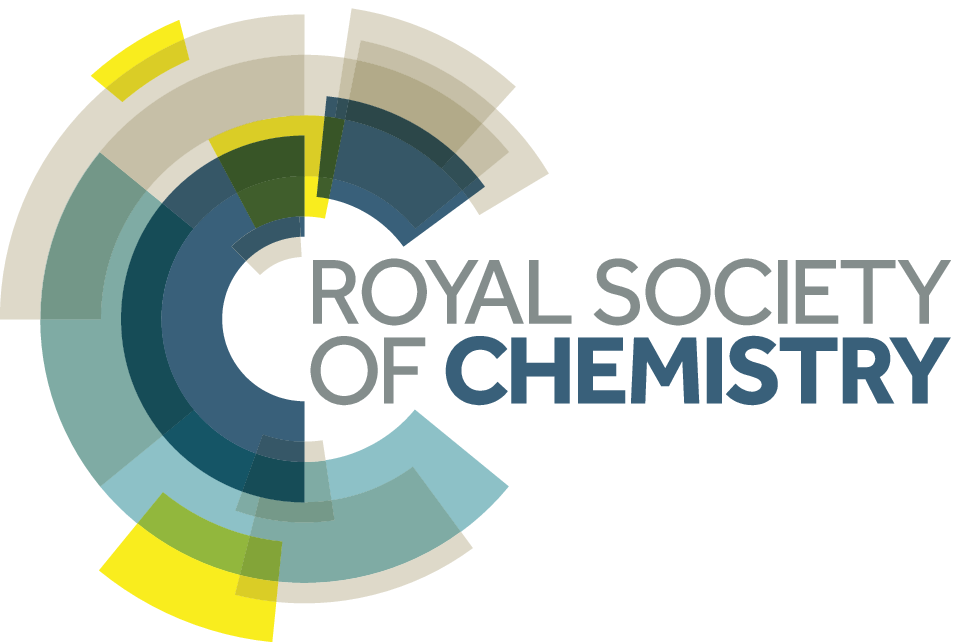}}
\renewcommand{\headrulewidth}{0pt}
}

\makeFNbottom
\makeatletter
\renewcommand\LARGE{\@setfontsize\LARGE{15pt}{17}}
\renewcommand\Large{\@setfontsize\Large{12pt}{14}}
\renewcommand\large{\@setfontsize\large{10pt}{12}}
\renewcommand\footnotesize{\@setfontsize\footnotesize{7pt}{10}}
\makeatother

\renewcommand{\thefootnote}{\fnsymbol{footnote}}
\renewcommand\footnoterule{\vspace*{1pt}%
\color{cream}\hrule width 3.5in height 0.4pt \color{black}\vspace*{5pt}} 
\setcounter{secnumdepth}{5}

\makeatletter 
\renewcommand\@biblabel[1]{#1}            
\renewcommand\@makefntext[1]%
{\noindent\makebox[0pt][r]{\@thefnmark\,}#1}
\makeatother 
\renewcommand{\figurename}{\small{Fig.}~}
\sectionfont{\sffamily\Large}
\subsectionfont{\normalsize}
\subsubsectionfont{\bf}
\setstretch{1.125} 
\setlength{\skip\footins}{0.8cm}
\setlength{\footnotesep}{0.25cm}
\setlength{\jot}{10pt}
\titlespacing*{\section}{0pt}{4pt}{4pt}
\titlespacing*{\subsection}{0pt}{15pt}{1pt}

\fancyfoot{}
\fancyfoot[LO,RE]{\vspace{-7.1pt}\includegraphics[height=9pt]{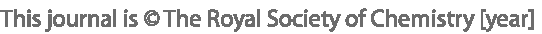}}
\fancyfoot[CO]{\vspace{-7.1pt}\hspace{13.2cm}\includegraphics{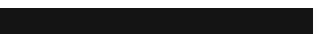}}
\fancyfoot[CE]{\vspace{-7.2pt}\hspace{-14.2cm}\includegraphics{head_foot/RF.eps}}
\fancyfoot[RO]{\footnotesize{\sffamily{1--\pageref{LastPage} ~\textbar  \hspace{2pt}\thepage}}}
\fancyfoot[LE]{\footnotesize{\sffamily{\thepage~\textbar\hspace{3.45cm} 1--\pageref{LastPage}}}}
\fancyhead{}
\renewcommand{\headrulewidth}{0pt} 
\renewcommand{\footrulewidth}{0pt}
\setlength{\arrayrulewidth}{1pt}
\setlength{\columnsep}{6.5mm}
\setlength\bibsep{1pt}

\makeatletter 
\newlength{\figrulesep} 
\setlength{\figrulesep}{0.5\textfloatsep} 

\newcommand{\topfigrule}{\vspace*{-1pt}%
\noindent{\color{cream}\rule[-\figrulesep]{\columnwidth}{1.5pt}} }

\newcommand{\botfigrule}{\vspace*{-2pt}%
\noindent{\color{cream}\rule[\figrulesep]{\columnwidth}{1.5pt}} }

\newcommand{\dblfigrule}{\vspace*{-1pt}%
\noindent{\color{cream}\rule[-\figrulesep]{\textwidth}{1.5pt}} }

\makeatother

\twocolumn[
  \begin{@twocolumnfalse}
\vspace{3cm}
\sffamily
\begin{tabular}{m{4.5cm} p{13.5cm} }

\includegraphics{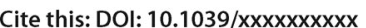} & \noindent\LARGE{\textbf{Vibrational and coherence dynamics in molecules$^\dag$}} \\
\vspace{0.3cm} & \vspace{0.3cm} \\

 & \noindent\large{Zhedong Zhang,\textit{$^{a}$} and Jin Wang$^{\ast,}$\textit{$^{a,}$}\textit{$^{b,}$}\textit{$^{c,}$}} \\

\includegraphics{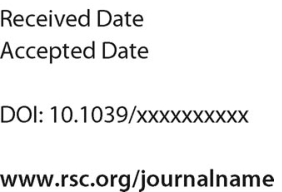} & \noindent\normalsize{We {\it analytically} investigate the population and coherence dynamics and relaxations in the vibrational energy transport in molecules. The corresponding two time scales $t_1$ and $t_2$ are explored. The coherence-population entanglement is found to considerably promote the time scale $t_2$ for dephasing and the amplitude of coherence. This is attributed to the suppression of the environment-induced drift force by the coherence. Moreover the population dynamics is shown to be significantly amplified and survive much longer with the coherence-population entanglement. Contrary to the previous studies, we exactly elucidate a coherent process by showing $t_1<t_2$. 
We predict the relaxation of vibrational and orientational dynamics of OH-stretching modes in consistence with the recent experiments, when applying to the water molecules dissolved in D$_{2}$O. Finally we explore the conherence effect on the heat current at macroscopic level.} \\

\end{tabular}

 \end{@twocolumnfalse} \vspace{0.6cm}

  ]

\renewcommand*\rmdefault{bch}\normalfont\upshape
\rmfamily
\section*{}
\vspace{-1cm}


\footnotetext{\textit{$^{a}$~Department of Physics and Astronomy, SUNY Stony Brook, Stony Brook, NY 11794, USA; Email: zhedong.zhang@stonybrook.edu }}
\footnotetext{\textit{$^{b}$~Department of Chemistry, SUNY Stony Brook, Stony Brook, NY 11794, USA; Tel: 1-631-632-1185; Fax: 1-631-632-7960; Email: jin.wang.1@stonybrook.edu }}
\footnotetext{\textit{$^{c}$~State Key Laboratory of Electroanalytical Chemistry, Changchun Institute of Applied Chemistry, Chinese Academy of Sciences, Changchun, Jilin 130022, P. R. China }}

\footnotetext{\dag~Electronic Supplementary Information (ESI) available: [details of any supplementary information available should be included here]. See DOI: }



\section{Introduction}
Vibrational energy transport, as one of the fundamental molecular motions, attracts much attention in recent studies of condensed phases, excitation energy transfer and chemical reaction energy dissipation \cite{Leitner12,Oxtoby81,Bakker93,Rubi10,Ishizaki10,Leitner13,Christensson12,Plenio13}. The significance of the vibrational energy transfer has been already recognized several decades ago \cite{Owrutsky94,Stratt95} and the classical kinetic description was well established in the condensed phase. However, the role of quantum effect, i.e., coherence and entanglement, in the intramolecular vibrational energy transport is still challenging \cite{Chen14,Bakker99}, especially in the far-from-equilibrium regime \cite{Wang08,Zhang14,Zhang15}.

Quantum nature, characterized by the coherence, has been explored as assisting the energy transfer in a wide range of complexes, such as the photosynthetic reaction center \cite{Engel07,Zhang15,Engel10,Scholes10,Plenio08} and the thermal transport in nano-device. Moreover, the coherence may also play a significant role in the intramolecular vibrational energy transfer, owing to the delocalization of the molecular stretching observed in experiments \cite{Bakker99}. The coherence relaxation was monitored in the experiments which uncovered the long-lived orientation dynamics with the presence of strong hydrogen bond between the D$_2$O molecules, accompanied with the resonant vibrational energy transfer in liquid water \cite{Bakker99,Bakker97}. 
The vibrational dynamics of molecules are usually characterized by two distinguished relaxation time scales \cite{Rice09,Miller91,Stanisz05,Harris91,Bonn07}, $t_1$ and $t_2$ associated with the population (longitudinal) and phase relaxations (decoherence), respectively. In the past studies it was always shown that $t_1\ge t_2$ \cite{Tyryshkin03,Silvestrini06,Clarke08}, which gives rise to the incoherent energy transport. However, there is a long debate on coherence assistance to the energy transfer at the molecular scale \cite{Kassal13,Fleming09,Fleming12}, that whether the long-lived coherence within $t_1<t_2$ considerably promotes the energy transfer in coherent regime. Many theoretical and experimental work addressed this issue recently and further provided some evidences of the oscillation feature of the coherence dynamics \cite{Engel10,Fleming07,Scholes12}.

In this article, we explore the vibrational and coherence dynamics in intermolecular vibrational energy transport in an analytical manner, by exploring the two typical time scales $t_1,\ t_2$. First we find that the phase relaxation is much slower than the population, namely, $t_1<t_2$. This, in other words, gives rise to the coherent energy transfer, in contrast to the incoherent one prediceted by F$\ddot{\textup{o}}$rster theory \cite{Forster46,Forster48,Schulten11}. Secondly the environment-induced coherence is found to considerably improve the phase-surviving time $t_2$ and amplify the amplitude of the coherence dynamics, by the comparison to the secular approximation where the coherence dynamics is decoupled to the populations. This result, on the other hand, can be applied to the processes (including both the bosonic and fermionic cases) described by quantum master equation in general \cite{Palmieri09}. By applying our model to the water molecules dissolved in D$_2$O, the prediction of typical time scale of orientational relaxation based on our theoretical investigation leads to the perfect agreement with the experiments \cite{Bakker99,Bakker97}. Moreover we also demonstrate that the coherence-population entanglement beyond the secular approximation has {\it non-trivial} contribution to the suppression of the energy consumption of the molecular mechine for reaching the nonequilibrium steady state. In other words, this indicates the enhancement of the energy transport efficiency.

\section{Two-oscillator model}
\subsection{Hamiltonian}
The molecular vibrations can be properly described by the two quantum-mechanically coupled oscillators with different excitation frequencies. In order to realize the vibrational energy transfer, the two-oscillator system needs to interact with two independent environments, one of which provides the excitation energy in molecules and the other harvests the dissipation energy. This model on the other hand, also captures some features of the energy transfer process in light-harvesting complex, even though it is the excitonic excitations in photosynthesis. The free and interaction Hamiltonian of system and environments read
\begin{equation}
\begin{split}
& H_0 = \varepsilon_1 a_1^{\dagger}a_1 + \varepsilon_2 a_2^{\dagger}a_2 + \Delta(a_1^{\dagger}a_2+a_2^{\dagger}a_1) + \sum_{\nu=1}^2\sum_{\textbf{k},\sigma}b_{\textbf{k}\sigma}^{(\nu),\dagger}b_{\textbf{k}\sigma}^{(\nu)}\\
& H_{int} = \sum_{\textbf{k},\sigma}g_{\textbf{k}\sigma}\left(c^{\dagger}b_{\textbf{k}\sigma}^{(1)} + c\  b_{\textbf{k}\sigma}^{(1),\dagger} \right) + \sum_{\textbf{q},s}f_{\textbf{q}s}\left(a_2^{\dagger}b_{\textbf{k}\sigma}^{(2)} + a_2 b_{\textbf{k}\sigma}^{(2),\dagger} \right)
\end{split}
\label{1}
\end{equation}
where $c\equiv a_1+a_2$ and $b_{\textbf{k}\sigma}^{(1)},\ b_{\textbf{k}\sigma}^{(2)}$ are the bosonic annihilation operators of environments. The rotating-wave approximation \cite{Scully97,Breuer02} has been applied to the vibration-bath interactions, owing to ignorance of virtual-process in long time limit. Notice that the dissipation environment only connects to one molecular vibration since in water molecules some of the surface OH groups couple to bulk water which is described by the harmonic oscillation bath \cite{Bonn11,Shen93,Shen06}. In realistic systems, the two vibrational modes do interact with other discrete vibrations, i.e., the stretching of OH bond in other molecules \cite{Shen06,Shen94}, which in some sense, can be treated as the vibron-phonon (VP) interaction. As pointed out previously \cite{Zhang15,Fassioli13}, the strong interactions between the system and some discrete vibrational modes due to the quasi-resonance between frequencies, leads to the comparable time scales of system and these vibrational modes, which subsequently acquires us to include the dynamics of these modes together with the system. In other words, these vibrational modes must be seperated from the bath degree of freedoms. They cause the renormalization of the coupling strength $\Delta$ between the excitations in system. The remaining modes consisting of low-energy fluctuations can then be resonably treated as the baths, which are in weak coupling to the systems owing to the mismatch of the frequencies between these continous modes and system.

\subsection{Quantum Master Equation}
Based on the perturbation theory with the rational given in last subsection, the whole solution of the density operator can be written as $\rho_{SR}=\rho_s(t)\otimes\rho_R(0)+\rho_{\delta}(t)$ with the traceless term in higher orders of system-bath coupling. From the Born-Markoff approximation where the time scale associated with the environmental correlations is much smaller than that of system over which the state varies appreciably, the reduced density matrix of the systems can be arrived by the operator master equation  $\dot{\rho}_s=\frac{i}{\hbar}[\rho_s,H_s]+\frac{1}{2\hbar^2}\mathcal{D}(\rho_s)$ with $H_s=\varepsilon_1 a_1^{\dagger}a_1 + \varepsilon_2 a_2^{\dagger}a_2 + \Delta(a_1^{\dagger}a_2+a_2^{\dagger}a_1)$ and 
\begin{equation}
\begin{split}
& \mathcal{D}(\rho_s) = \sum_{j,p=1}^2\left[\gamma_p^{T_1,+}\left(a_p\rho_s a_j^{\dagger}-a_j^{\dagger}a_p\rho_s\right)+\gamma_p^{T_1,-}\left(a_p^{\dagger}\rho_s a_j-a_ja_p^{\dagger}\rho_s\right)\right]\\
& \quad +\sum_{p=1}^2\left[\gamma_p^{T_2,+}\left(a_p\rho_s a_2^{\dagger}-a_2^{\dagger}a_p\rho_s\right)+\gamma_p^{T_2,-}\left(a_p^{\dagger}\rho_s a_2-a_2 a_p^{\dagger}\rho_s\right)\right] + \textup{h.c.}
\end{split}
\label{2}
\end{equation}
where the reservoirs are assumed to be in thermal equilibrium. The expressions of dissipation rates $\gamma_p^{T_{\nu},\pm}$ will be given in Supplementary Information (SI) and $T_{\nu}$'s are the temperatures of environments. 


We will solve the QME in coherent representation which differs from the conventional way in Liouville space. This method was first developed by Glauber \cite{Glauber63}. It is alternatively named as Glauber-Sudarshan $P$ representation in quantum optics. As is known the density matrix is expanded in terms of the eigenstates of the annihilation operators \cite{Carmichael99}
\begin{equation}
\begin{split}
\rho_s(t) = \int P(\alpha_{\mu},\alpha_{\mu}^*,t)|\alpha_1,\alpha_2\rangle\langle\alpha_1,\alpha_2|d^2\alpha_1d^2\alpha_2
\end{split}
\label{5}
\end{equation}
where $\hat{a}_j|\alpha_1,\alpha_2\rangle=\alpha_j|\alpha_1,\alpha_2\rangle$ and $P(\alpha_{\mu},\alpha_{\mu}^*,t)$ is called quasi-probability, due to the overcompleteness of the coherent basis. By projecting into the coherent representation, the QME is in the form of PDE

\begin{widetext}
\begin{equation}
\begin{split} \label{6}
\frac{\partial}{\partial t}P(\alpha_{\mu},\alpha_{\mu}^*,t) =  \bigg[\left(i\omega_1+\gamma\right)\frac{\partial}{\partial\alpha_1}\alpha_1 & +\left(i\omega_2+2\gamma\right)\frac{\partial}{\partial\alpha_2}\alpha_2 +\left(iu+\epsilon\gamma\right)\left(\frac{\partial}{\partial\alpha_1}\alpha_2+\frac{\partial}{\partial\alpha_2}\alpha_1\right)+\textup{c.c.}\bigg]P(\alpha_{\mu},\alpha_{\mu}^*,t)\\[0.06cm] & \qquad\qquad + \gamma\bigg[2\textup{Y}_1^1\frac{\partial^2}{\partial\alpha_1^*\partial\alpha_1}+2\textup{Y}_2^2\frac{\partial^2}{\partial\alpha_2^*\partial\alpha_2} +\epsilon\textup{Y}_{12}^{21}\left(\frac{\partial^2}{\partial\alpha_1^*\alpha_2}+\frac{\partial^2}{\partial\alpha_1\partial\alpha_2^*}\right)\bigg]P(\alpha_{\mu},\alpha_{\mu}^*,t)
\end{split}
\end{equation}
\end{widetext}
\noindent with $\omega_j=\varepsilon_j/\hbar,\ u=\Delta/\hbar$. $\gamma=\pi D(\bar{\nu})g_{\bar{\nu}}^2/\hbar^2$ and $D(\varepsilon)$ is the density of states (DOS) which is a smooth function. The coupling between the coherence and population dynamics is governed by the adiabatic parameter $\epsilon$, of which the importance will be uncovered in Sec.4. To solve the dynamical equation Eq.(\ref{6}) we adopt the approach illustrated in the literature for Ornstein-Uhlenbeck process \cite{Risken89} and then write down the drift as well as diffusion matrices
\begin{equation}
\begin{split}
   \Sigma = \begin{pmatrix}
             \Gamma & 0\\[0.1cm]
             0 & \Gamma^{\dagger}\\
            \end{pmatrix},\ \
   D = \gamma\begin{pmatrix}
              0 & \textup{M}\\[0.1cm]
              \textup{M} & 0\\
             \end{pmatrix}
\end{split}
\label{sig}
\end{equation}
where
\begin{equation}
\begin{split}
\Gamma = \begin{pmatrix}
             i\omega_1+\gamma & iu+\epsilon\gamma\\[0.1cm]
             iu+\epsilon\gamma & i\omega_2+2\gamma\\
            \end{pmatrix},\ \
\textup{M} = \begin{pmatrix}
              \textup{Y}_1^1 & \frac{\epsilon}{2}\textup{Y}_{12}^{21}\\[0.1cm]
              \frac{\epsilon}{2}\textup{Y}_{12}^{21} & \textup{Y}_2^2\\
             \end{pmatrix}
\end{split}
\label{9}
\end{equation}
To solve the PDE in Eq.(\ref{6}) above, we need to get the eigenvalues and biorthogonal eigenvectors of the drift matrix, which will be shown in detail in SI. Here two quantities $F$ and $G$ are introduced
\begin{equation}
\begin{split}
& F = \sqrt{\frac{1}{2}\bigg[1+4\epsilon-4d^2-w^2+\sqrt{(1+4\epsilon-4d^2-w^2)^2+4(w-4\epsilon d)^2}\bigg]}\\
& G = \frac{4\epsilon d-w}{F},\ F^2+G^2 = \sqrt{(1+4\epsilon-4d^2-w^2)^2+4(w-4\epsilon d)^2}
\end{split}
\label{11}
\end{equation}
where $p_{\pm}=1\pm F,\ q_{\pm}=G\mp w$, $d=\frac{\Delta}{\hbar\gamma}$ and $w=\frac{\omega_1-\omega_2}{\gamma}$. Initially the system is properly assumed to stay at the ground state $\rho_0=|0,0\rangle\langle 0,0|$, since there is no excitation at the beginning. To solve the PDE above, one needs to obtain the Glauber representation of the initial state $\rho_0$. First we get the matrix element
$\langle -\alpha_1,-\alpha_2|\rho_0|\alpha_1,\alpha_2\rangle = e^{-(|\alpha_1|^2+|\alpha_2|^2)}$, 
which leads to the Glauber representation of the initial state based on the Fourier transform in the complex domain
\begin{equation}
\begin{split}
P(\alpha_{\mu},\alpha_{\mu}^*,0) & = \frac{e^{|\alpha_1|^2+|\alpha_2|^2}}{\pi^4}\int\int d^2\beta_1d^2\beta_2\langle -\beta_1,-\beta_2|\rho_0|\beta_1,\beta_2\rangle\\
& \times e^{|\beta_1|^2+|\beta_2|^2} e^{2i\textup{Im}(\beta_1^*\alpha_1+\beta_2^*\alpha_2)} = \delta^{(2)}(\alpha_1)\delta^{(2)}(\alpha_2)
\end{split}
\label{coh}
\end{equation}

\noindent Notice that the measure we used is $d^2\alpha=\textup{d}(\textup{Re}\alpha)\textup{d}(\textup{Im}\alpha)$. Therefore under the initial condition (\ref{coh}), the full solution to the dynamical equation Eq.(\ref{6}) is $P(\alpha_{\mu},\alpha_{\mu}^*,t)=\frac{a(t)b(t)-|c(t)|^2}{\pi^2}\textup{exp}\{-[a(t)|\alpha_1|^2+b(t)|\alpha_2|^2+c(t)\alpha_1^*\alpha_2+c^*(t)\alpha_1\alpha_2^*]\}$ and
\begin{equation}
\begin{split} & a(t)=\frac{\textup{A}_{11}^{24}\textup{Y}_1^1+\textup{A}_{22}^{24}\textup{Y}_2^2+\textup{A}_{1221}^{24}\textup{Y}_{12}^{21}}{\textup{det}(B)}\\ & b(t)=\frac{\textup{A}_{11}^{13}\textup{Y}_1^1+\textup{A}_{22}^{13}\textup{Y}_2^2+\textup{A}_{1221}^{13}\textup{Y}_{12}^{21}}{\textup{det}(B)}\\ & c(t)=-\frac{\textup{A}_{11}^{14}\textup{Y}_1^1+\textup{A}_{22}^{14}\textup{Y}_2^2+\textup{A}_{1221}^{14}\textup{Y}_{12}^{21}}{\textup{det}(B)}
\end{split}
\label{12}
\end{equation}
The coefficients $\textup{A}_{...}^{...}$ are given in SI.

\section{Coherence and population dynamics}
Given a density matrix representing the state of the molecular vibrations, we wish to evaluate the amount of entanglement in the state, which refers to non-local correlations between the vibrational modes of spatially seperated molecules. The mixed-state entanglement entropy quantifying the degree of entanglement of mixture ensemble is still an open question, despite the fact that it is well defined for the pure state. Another measure of entanglement is the concurrence, which is computable for only two qubits. Here we choose the coherence
\begin{equation}
\begin{split}
C[\rho] = \textup{Tr}(\rho_sa_1^{\dagger}a_2)=\sum_{n_1=1}^{\infty}\sum_{n_2=1}^{\infty}\sqrt{n_1n_2}\ \langle n_1-1,n_2|\rho_s|n_1,n_2-1\rangle
\end{split}
\label{14}
\end{equation}
to quantify the entanglement between different vibrational modes, from the combination of off-diagonal elements of density matrix in Fock space. First this quantity is basis-independent while the conventional description is not. Secondly as reflected in the operator master equation Eq.(\ref{2}) the $C[\rho]$ introduced here interacts with the populations, which in other words, may have significant contribution to the population dynamics. In analogy with the NMR experiment, the population dynamics is governed by the polarization
\begin{equation}
\begin{split}
M_z = \frac{n_1-n_2}{n_1+n_2}
\end{split}
\label{15}
\end{equation}
or alternatively $M_z$ can be also called population imbalance. $n_1=\textup{Tr}(\rho_sa_1^{\dagger}a_1),\ n_2=\textup{Tr}(\rho_sa_2^{\dagger}a_2)$. In our model, these two observables are written as
\begin{equation}
\begin{split}
& C[\rho] = \textup{A}_{11}^{14,*}\textup{Y}_1^1+\textup{A}_{22}^{14,*}\textup{Y}_2^2+\textup{A}_{1221}^{14,*}\textup{Y}_{12}^{21}\\
& M_z = \frac{(\textup{A}_{11}^{13}-\textup{A}_{11}^{24})\textup{Y}_1^1+(\textup{A}_{22}^{13}-\textup{A}_{22}^{24})\textup{Y}_2^2+(\textup{A}_{1221}^{13}-\textup{A}_{1221}^{24})\textup{Y}_{12}^{21}}{(\textup{A}_{11}^{13}+\textup{A}_{11}^{24})\textup{Y}_1^1+(\textup{A}_{22}^{13}+\textup{A}_{22}^{24})\textup{Y}_2^2+(\textup{A}_{1221}^{13}+\textup{A}_{1221}^{24})\textup{Y}_{12}^{21}}
\end{split}
\label{16}
\end{equation}
The first column of Fig.\ref{f1} illustrates the coherence dynamics. It is commonly shown that the coherence is generated by the environments, before the dephasing takes place. Fig.\ref{f1}(a,b) show that (i) the sharp increase of both coherence as well as population imbalance and (ii) the considerable promotion of the amplitude of both coherence as well as population imbalance contributed from the thermal fluctuations in environments. In other words, these results elucidate that the environments do not only cause the dephasing process, but can also considerably enhance the quantum coherence, especially at the beginning of the dynamics. In fact  the origination of (ii) is from the improvement of steady-state coherence in far-from-equilibrium regime, as uncovered before \cite{Zhang14,Zhang15njp}.

It is widely known that the intramolecular relaxation process is governed by two distinct time scale $t_1$ and $t_2$ where $t_1$ refers to the time constant of longitudinal relaxation or relaxation in z-direction and $t_2$ refers to the transverse relaxation or phase relaxation. Physically $t_1$ relaxation describes the process of re-establishing the normal Guassian distribution of populations in states in the presence of environments. $t_2$ is the loss of phase correlation among molecules. Classically $t_1\ge t_2$. In quantum systems, however as shown later, $t_2$ can be larger than $t_1$, which indicates a strong coherent nature and further means that the coherence will survive during the process of energy or charge transfer. Roughly by comparing the decay tails between the two columns in Fig.\ref{f1} it seems to be apparent that the relaxation of phase coherence is slower than the longitudinal relaxation. This is reflected by the tail of decay which is smooth for coherence while it is of sharp decrease for population. To quantify this issue in detail, we need to estimate the time length of relaxation by $e^{-1}$ decay, since the behavior of time evolutions of population and coherence is of exponential feature as reflected in $\textup{A}_{...}^{...}$ and Eq.(\ref{16}).
\begin{table}
\centering
\begin{tabular}{ccccccc}
\hline\hline
$T_1$(K) & $\qquad$ & $t_1$(fs) & $\qquad$ & $t_2$(fs) & $\qquad$ & $t_2/t_1$\\[0.5ex]
\hline
3500 & $\qquad$ & 36 & $\qquad$ & 71.6 & $\qquad$ & 1.98\\
5600 & $\qquad$ & 35.8 & $\qquad$ & 70.4 & $\qquad$ & 1.97\\
8000 & $\qquad$ & 35.6 & $\qquad$ & 80 & $\qquad$ & 2.25\\[1ex]
\hline
\end{tabular}
\caption{Time constants $t_1$ and $t_2$ with different temperatures. $\Delta=0.1$eV, $\delta\varepsilon=0.15$eV, $T_2=2100$K and $\gamma=10$ps$^{-1}$.}
\label{timetemp}
\end{table}
Both Table I and II show that the longitudinal relaxation is faster than the coherence relaxation, namely, $t_1<t_2$ which is contrary to the usual cases \cite{Silvestrini06,Clarke08}. On the other hand, both of longitudinal and coherence relaxations are nearly not affected by the thermal fluctuations in reservoirs while they are sensitive to the vibron-vibron interaction. This is because the $F,\ G$ defined before are independent of the bath parameters. In particular, the longitudinal relaxation becomes faster while the coherence relaxation becomes slower, as the vibron-vibron coupling increases. This originates from the vibron-phonon interaction (i.e., hydrogen bond) which leads to the renormalization of vibron-vibron couplings. Furthermore, the slower relaxation of phase coherence indicates that the dynamical energy transport process is more coherent at large vibron-vibron couplings.
\begin{table}
\centering
\begin{tabular}{ccccccc}
\hline\hline
$\Delta$(eV) & $\qquad$ & $t_1$(fs) & $\qquad$ & $t_2$(fs) & $\qquad$ & $t_2/t_1$\\[0.5ex]
\hline
0.005 & $\qquad$ & $--$ & $\qquad$ & 14.6 & $\qquad$ & $--$\\
0.08  & $\qquad$ & 28.3 & $\qquad$ & 79.1 & $\qquad$ & 2.8\\
0.3   & $\qquad$ & 28.3 & $\qquad$ & 112 & $\qquad$ & 3.96\\[1ex]
\hline
\end{tabular}
\caption{Time constants $t_1$ and $t_2$ with various vibron-vibron couplings. $\delta\varepsilon=0.15$eV, $T_1=5600$K, $T_2=2100$K and $\gamma=10$ps$^{-1}$.}
\label{timedelta}
\end{table}

By applying our model to the OH-stretching mode of HDO dissolved in D$_2$O, the parameters are $\varepsilon_1=3500$cm$^{-1}$, $\varepsilon_2=3320$cm$^{-1}$ and $\gamma=0.3$ps$^{-1}$ according to Ref. \cite{Bakker97}. The vibrational and orientational dynamics of OH-stretching mode of HDO molecules was measured by the femtosecond mid-infrared pump-probe study. Here, the orientational dynamics refers to the coherence dynamics in our terminology. Our theoretical investigation illustrates the coherence (orientation) dynamics in Fig.\ref{f2}(left). 

The $e^{-1}$-decay-estimation of the time constant for coherence-surviving gives $\tau\simeq 2.6$ps associated with strongly hydrogen-bonded water molecules, whereas $\tau\simeq 0.7$ps for the weakly hydrogen-bonded water molecules. These are in good agreement (at least qualitatively, except the minor deviation caused by the simplification of theoretical model) with the measurements in recent experiments \cite{Bakker99,Bakker97}, where the measurement of the time constants gives $\tau\simeq 13$ps and $0.7$ps, respectively.

\section{Quantification of coherence effect}
\begin{figure}
\centering
 $\begin{array}{cc}
  \includegraphics[scale=0.325]{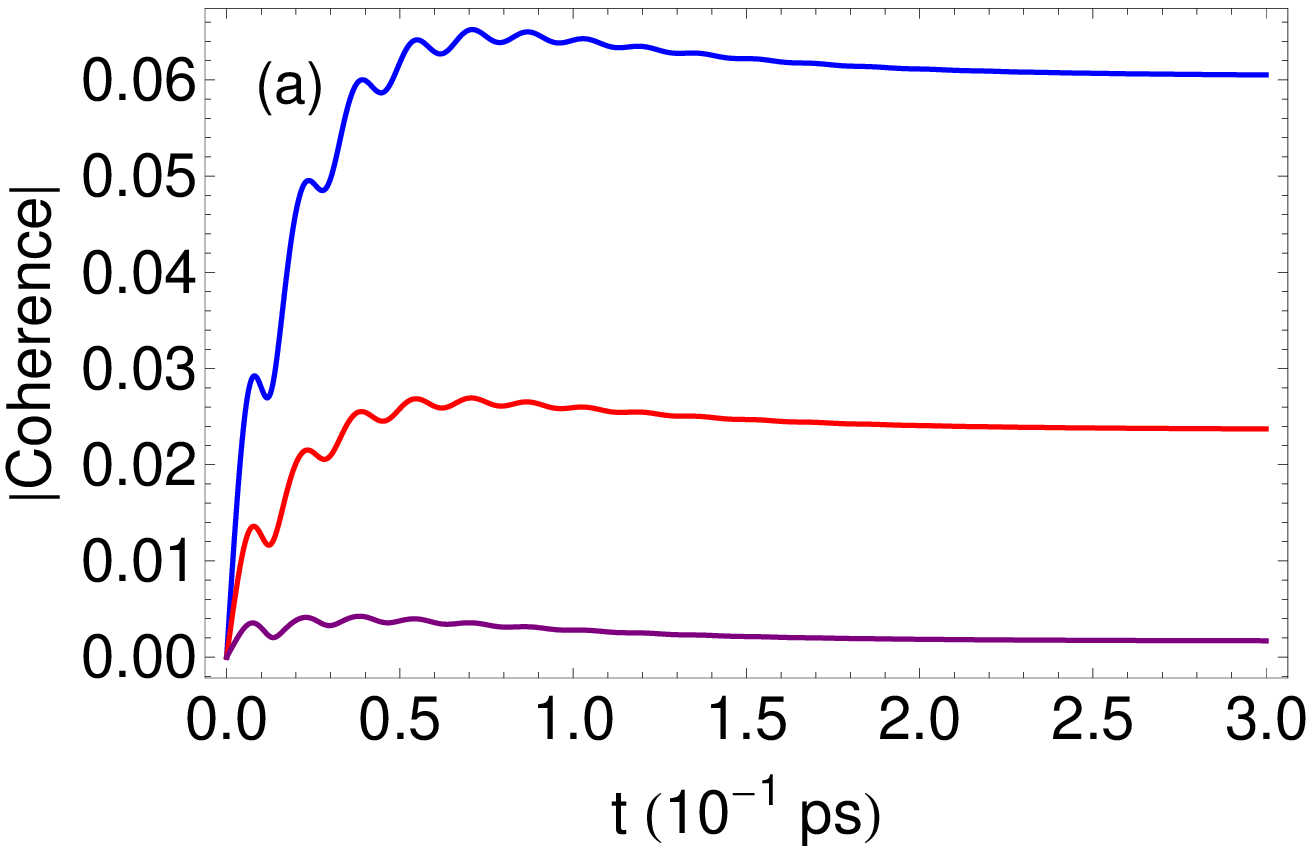}
 &\includegraphics[scale=0.315]{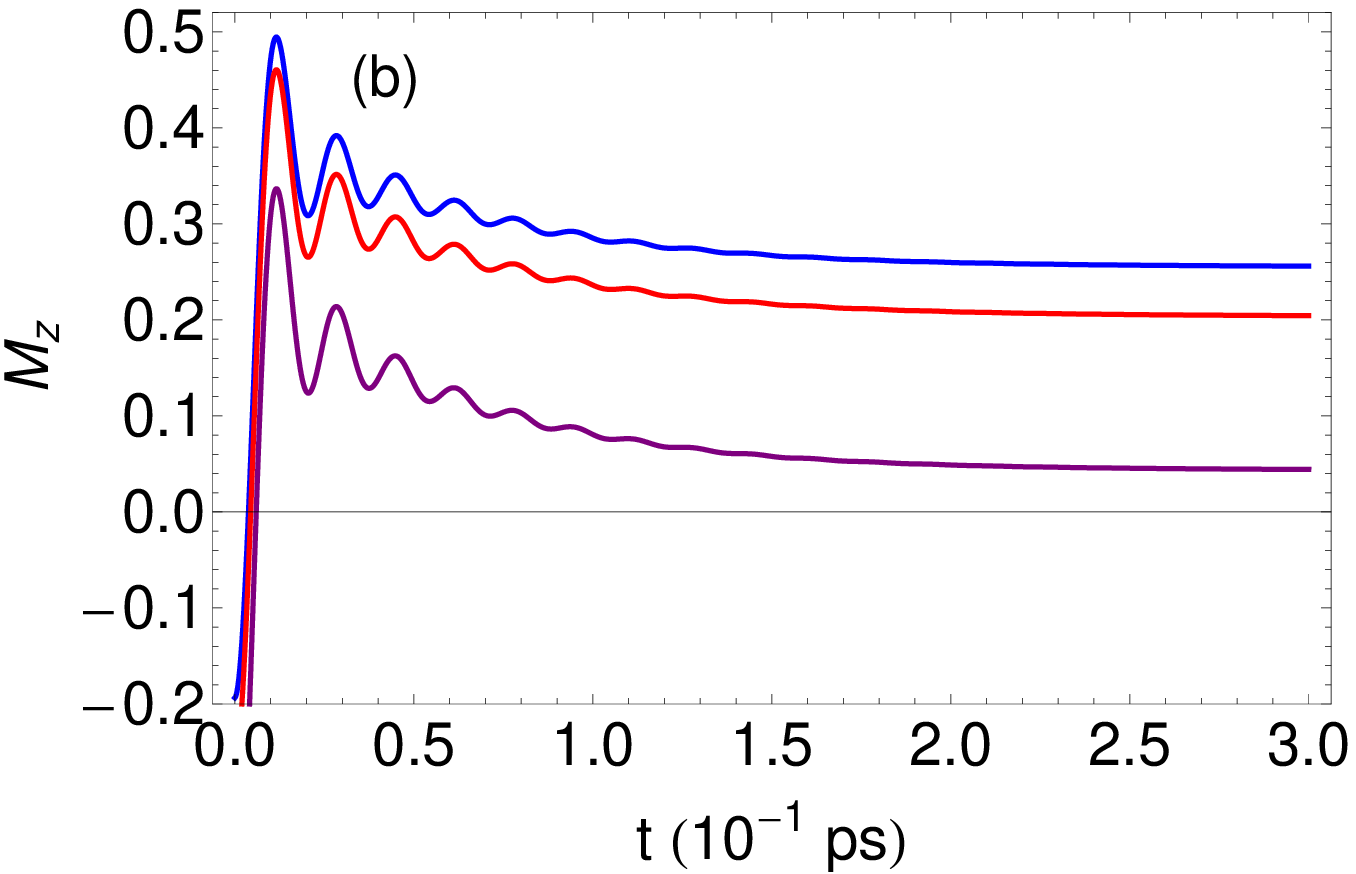}\\
  \includegraphics[scale=0.325]{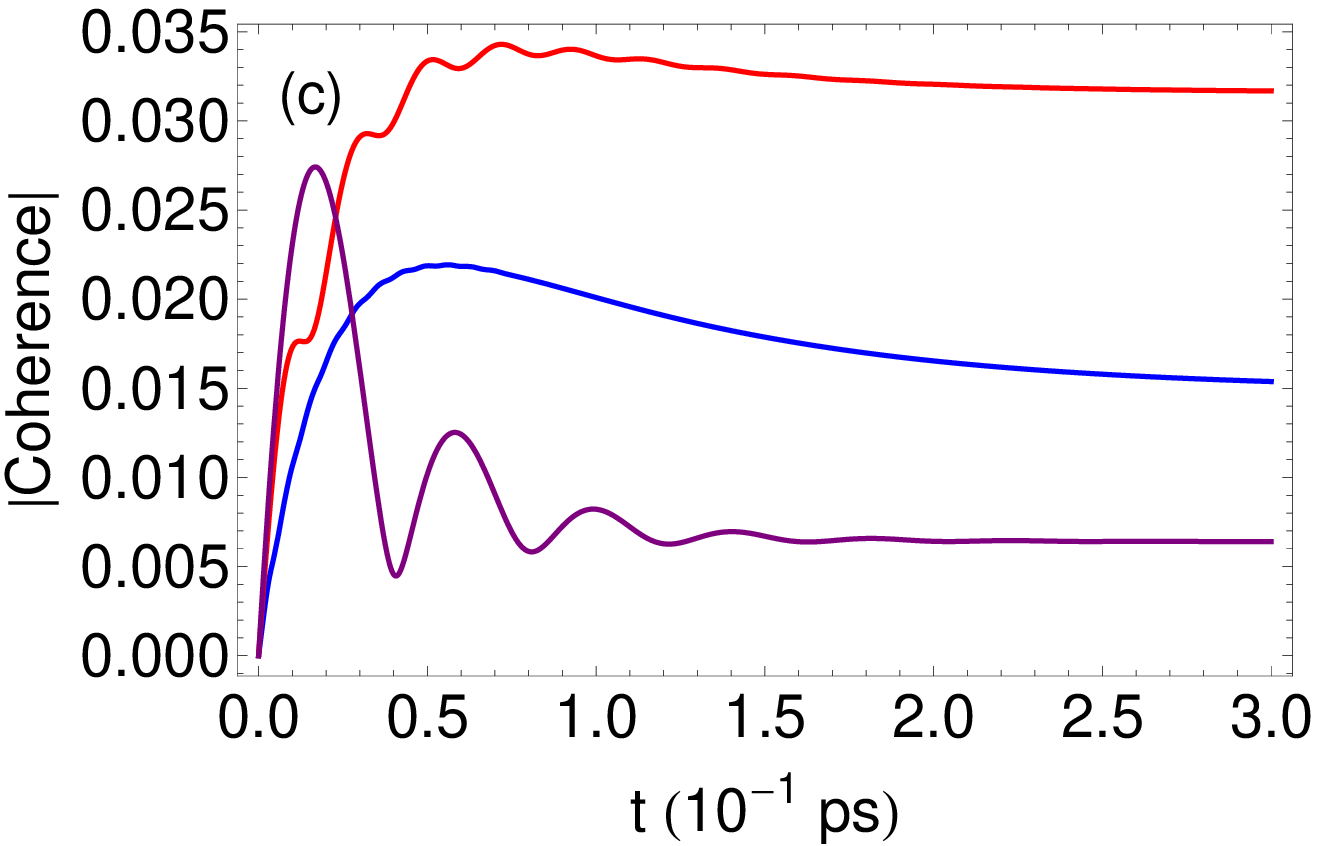}
 &\includegraphics[scale=0.325]{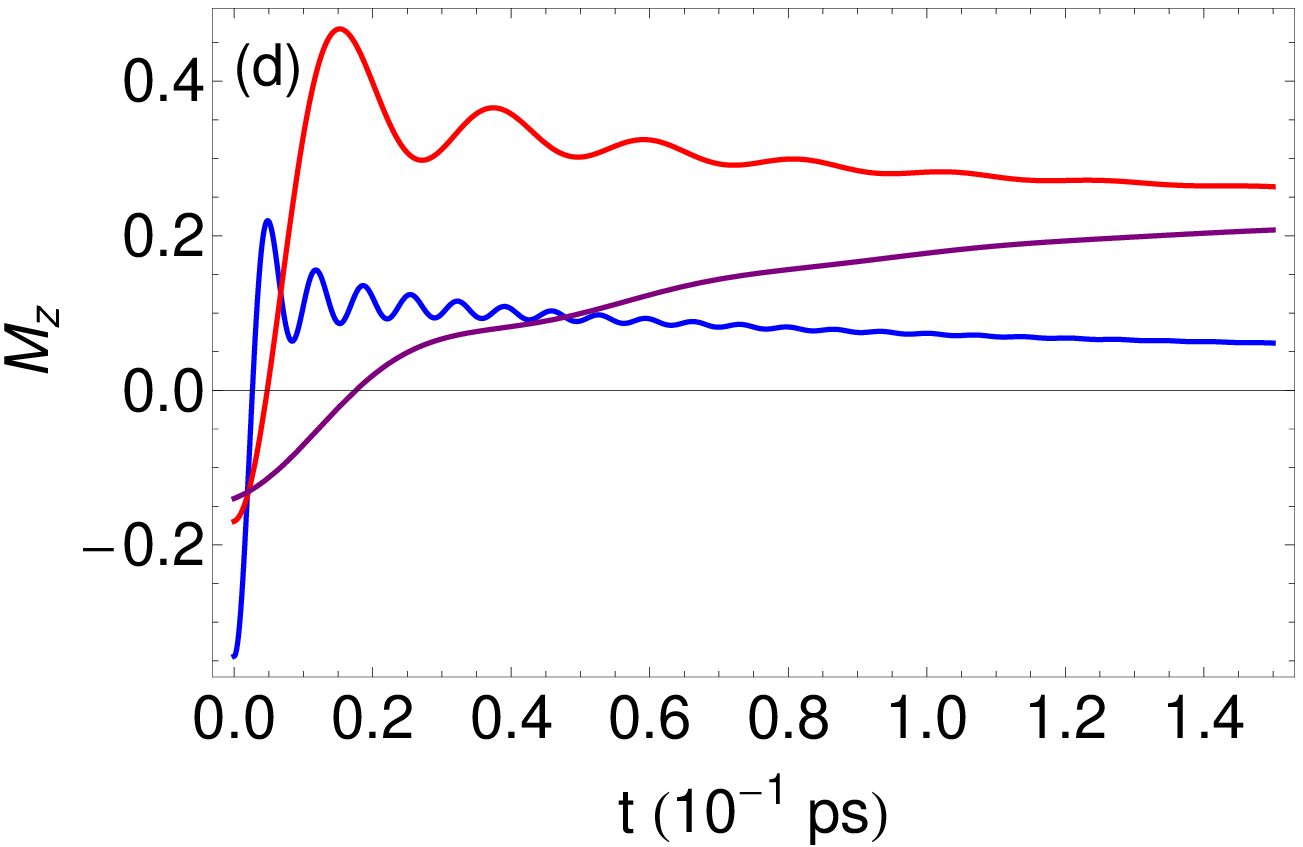}\\
  \includegraphics[scale=0.325]{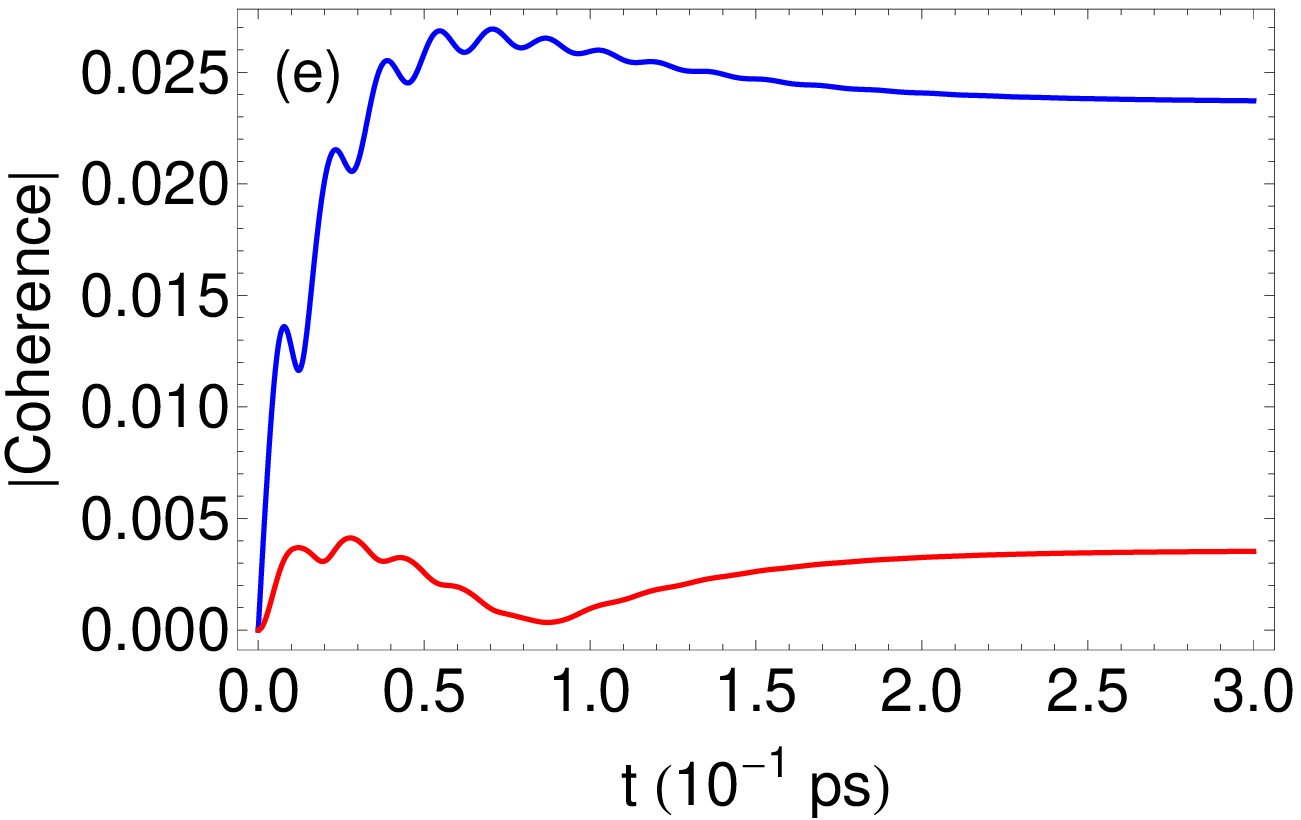}
 &\includegraphics[scale=0.325]{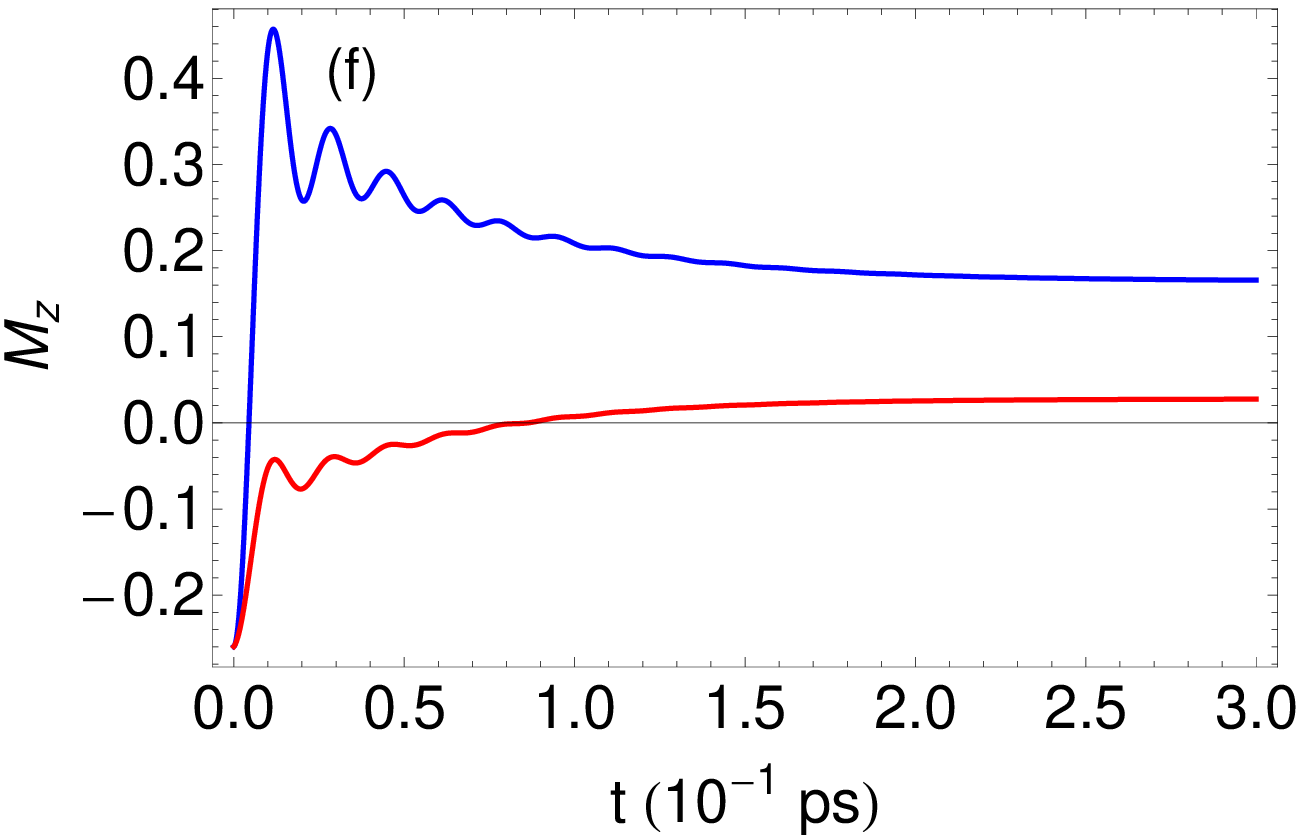}
  \end{array}$
\caption{(Color online) Time evolution of coherence and magnetization (population imbalance) under various (a,b) $T_1$ measuring the thermal fluctuations of environments, (c,d) vibron-vibron interactions and (e,f) coherence-population entanglement; In (a,b) the blue, red and purple lines are for $T_1=8000$K, $5600$K and $3500$K, respectively. $\Delta=0.1$eV; In (c,d) the blue, red and purple lines are for $\Delta=0.3$eV, $0.08$eV and $5$meV, respectively. $T_1=5600$K; In (e,f) the blue and red curves correspond to the cases without and with secular approximation. Other paramters are $\delta\varepsilon=0.15$eV, $T_2=2100$K and $\gamma=10$ps$^{-1}$.}
\label{f1}
\end{figure}

There was long debate on the coherence contribution to the vibrational and coherence dynamics, and also the energy (charge) transport. By introducing an adiabatic parameter $\epsilon$ in Eq.(\ref{2}) and (\ref{6}) before, here we will study how the coherence-population entanglement gradually affects the relaxation process. Notice that the coherence-population entanglement is mainly generated by the environments. We will further perform a comparison between our full quantum dynamics ($\epsilon=1$) and the one within the secular approximation ($\epsilon=0$), in which the coherence and population dynamics are unentangled. The secular approximation has been popularly appiled to the Lindblad equation describing the chemical reactions and light-harvesting complex, so that its validity should be examined hereafter. The dissipation term in Eq.(\ref{2}) reduces to
\begin{equation}
\begin{split}
& \bar{\mathcal{D}}(\rho_s) = \sum_{j=1}^2 \left[\gamma_j^{T_1,+}\left(a_j\rho_s a_j^{\dagger}-a_j^{\dagger}a_j\rho_s\right)+\gamma_j^{T_1,-}\left(a_j^{\dagger}\rho_s a_j-a_ja_j^{\dagger}\rho_s\right)\right]\\
& \qquad + \gamma_2^{T_2,+}\left(a_2\rho_sa_2^{\dagger}-a_2^{\dagger}a_2\rho_s\right)+\gamma_2^{T_2,-}\left(a_2^{\dagger}\rho_sa_2-a_2a_2^{\dagger}\rho_s\right)+\textup{h.c.}
\end{split}
\label{17}
\end{equation}
which leads to the case $\epsilon=0$ in Eq.(\ref{6}). Correspondingly the solution shares the same expression as Eq.(\ref{6}) by setting $\epsilon=0$. Therefore the population imbalance (magnetization) and coherence are in the same formalisms as the ones in previous Eq.(\ref{16}), under the limit $\epsilon\rightarrow 0^+$. Moreover the residue coherence in long time limit even under secular approximation (shown in Fig.\ref{f1}(e)) is due to the coupling between vibrational modes (vibron-vibron interaction as mentioned before), although in the dissipation part the coherence is decoupled from population dynamics.

\begin{figure}
\centering
 $\begin{array}{cc}
  \includegraphics[scale=0.32]{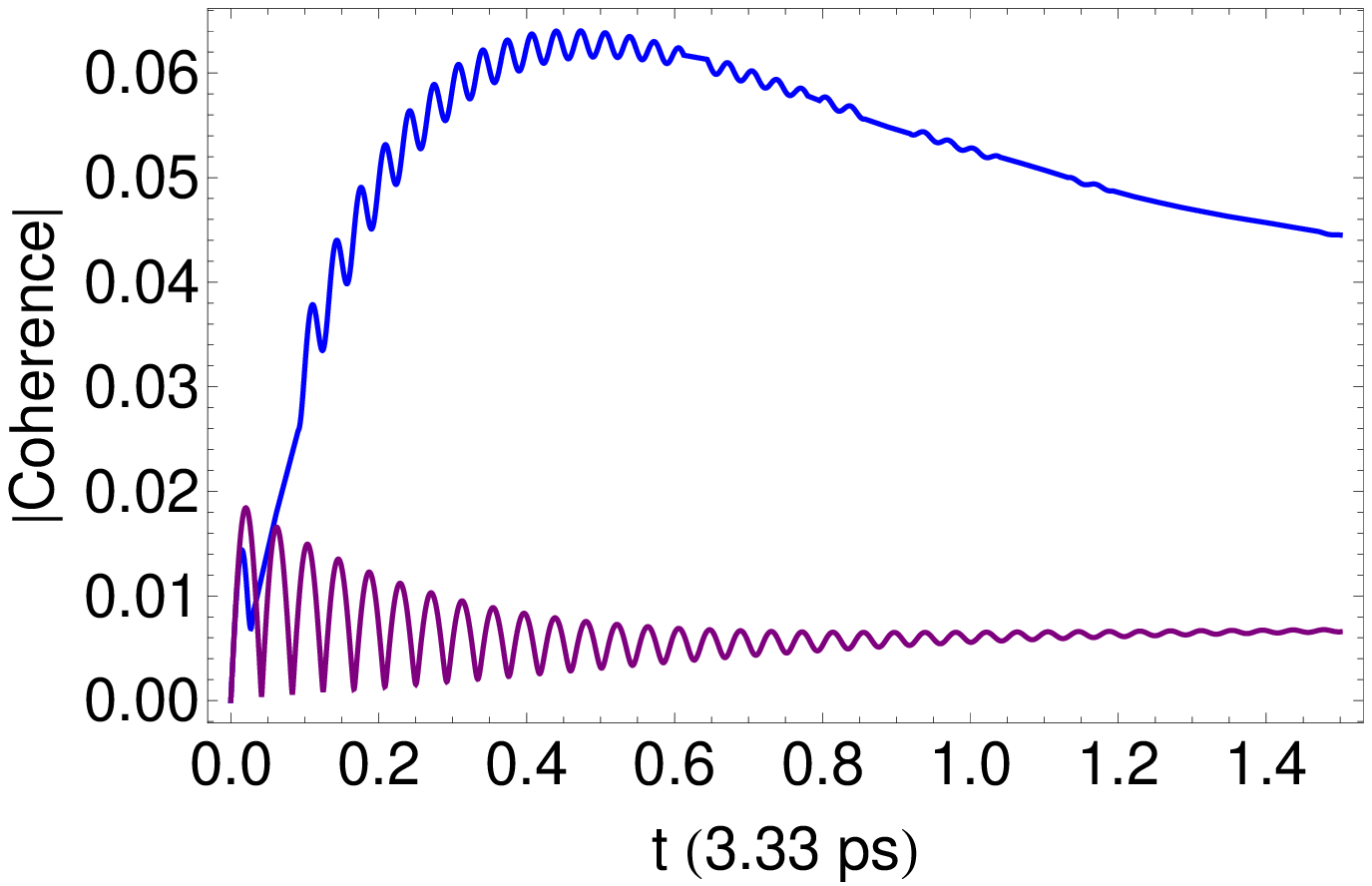}\ \ 
  \includegraphics[scale=0.4]{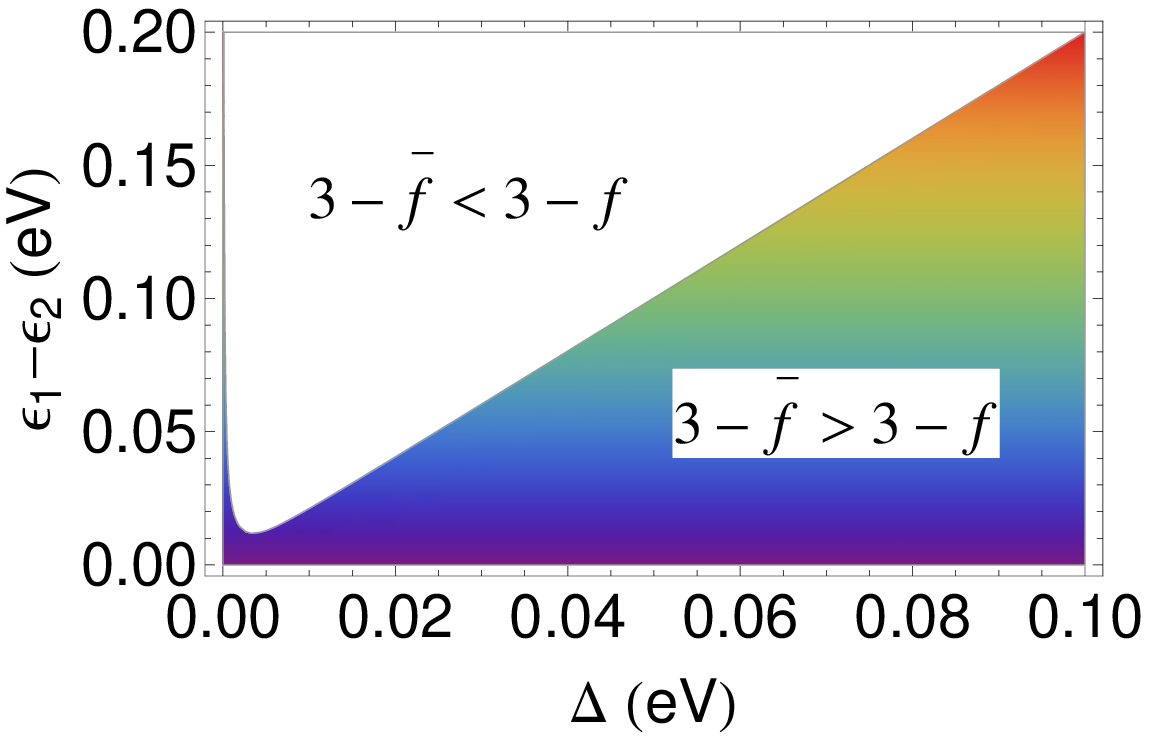}
  \end{array}$
\caption{(Color online) (Left) Orientation (coherence) dynamics for the OH-stretching mode of HDO dissolved in D$_2$O and (right) 2D plot of $\frac{3-F|_{\epsilon=0}}{3-F|_{\epsilon=1}}$ as a function of $\Delta$ and $\delta\varepsilon$. (Left) The blue and purple lines correspond to $\Delta=0.0112$eV and $0.5$meV, respectively. Other parameters are $T_1=5600$K and $T_2=2100$K}
\label{f2}
\end{figure}

\begin{figure}
\centering
 $\begin{array}{cc}
  \includegraphics[scale=0.35]{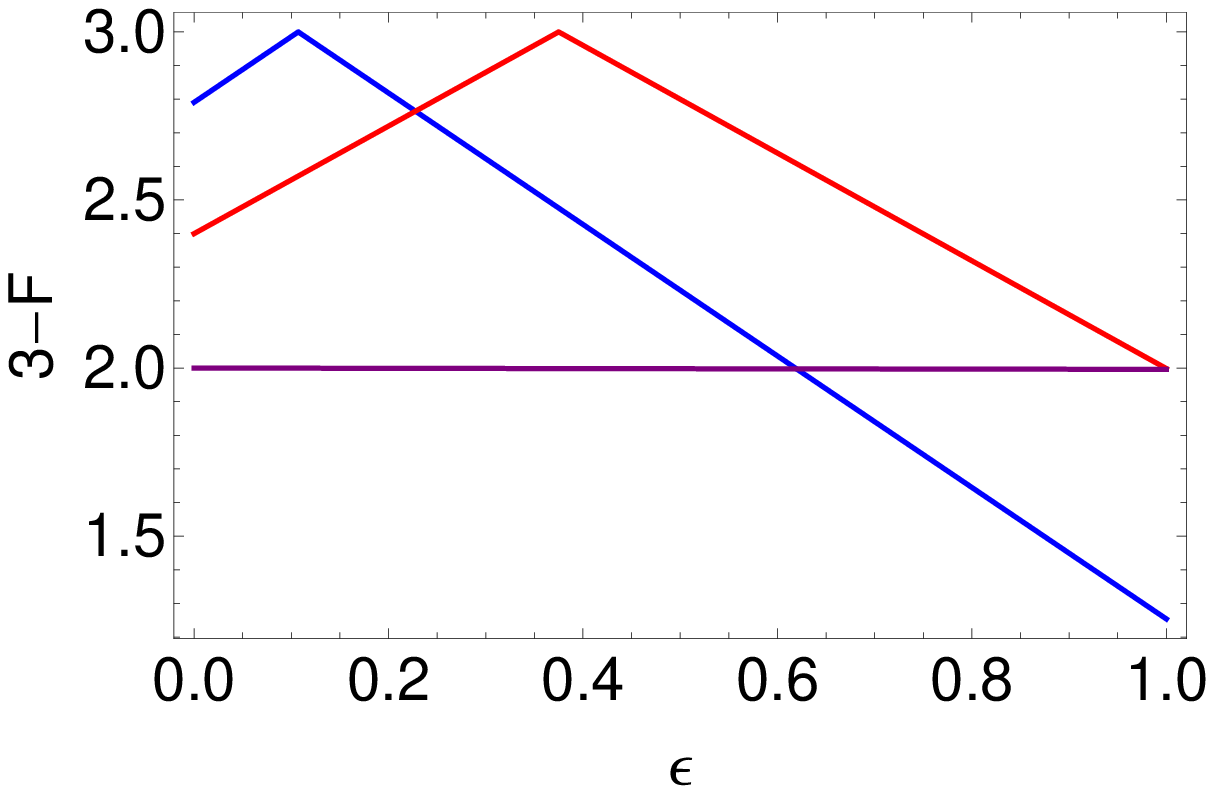}\ 
  \includegraphics[scale=0.32]{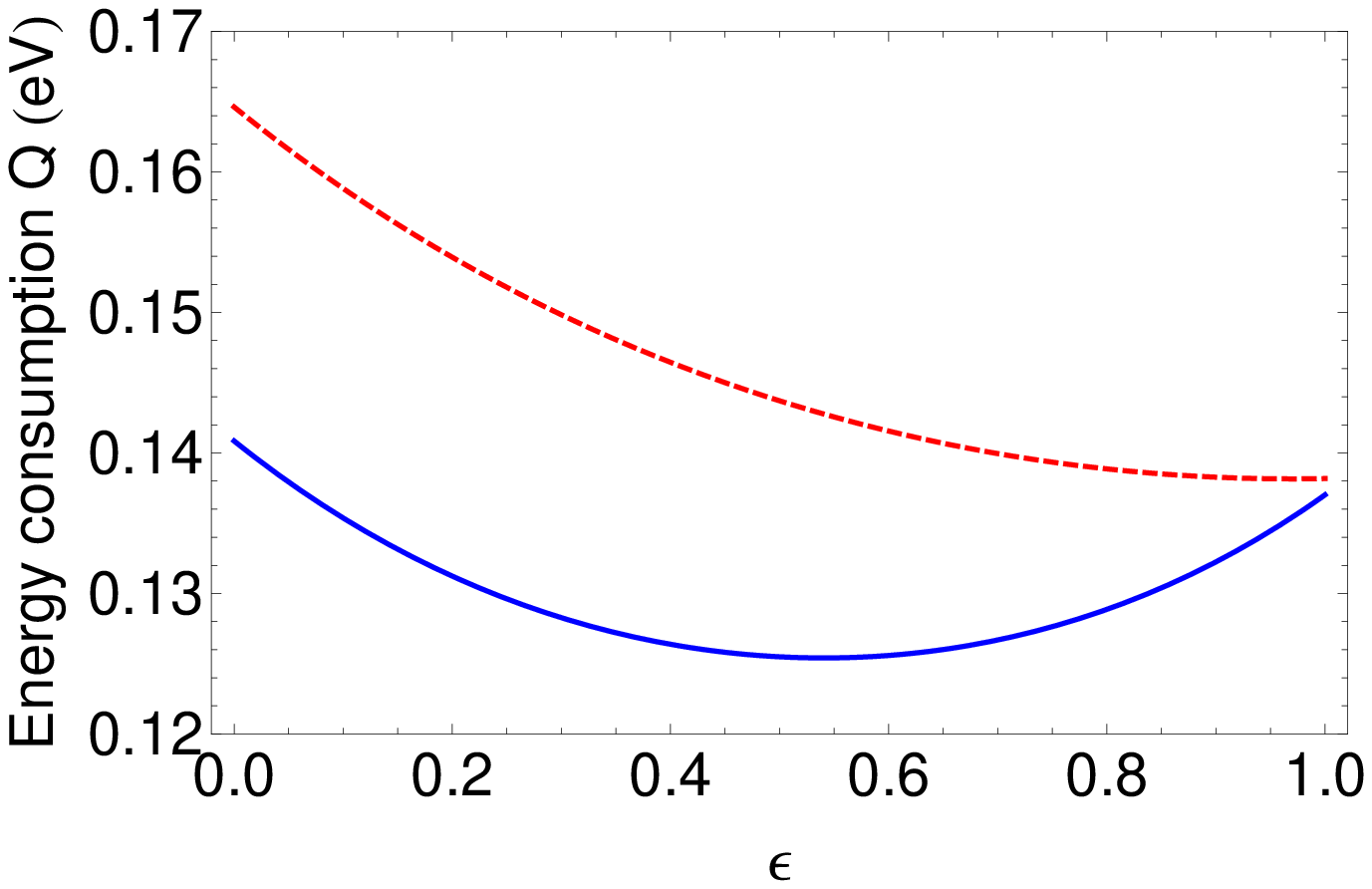}
  \end{array}$
\caption{(Color online) (Left) The parameter $3-F$ varies as a function of $\epsilon$. The blue, red and purple lines are for $\Delta=0.35$eV, $0.1$eV and $0$, respectively; (Right) Energy consumption as a function of $\epsilon$, where blue and red lines are for $\Delta=0.1$eV and $0.05$eV, respectively. Other parameters are $\delta\varepsilon=0.15$eV and $\gamma=10$ps$^{-1}$.}
\label{F}
\end{figure}

As is shown in Fig.\ref{f1}(e) and \ref{f1}(f), for phase coherence we found that (i) the amplitude is considerably improved and (ii) the time constant for decay is significantly extended (see the tail of decay), by the coherence-population entanglement. Quantitative analysis gives rise to $t_2\simeq 81$fs and $37.3$fs for including coherence-population entanglement and not, respectively. This behavior is attributed to the reduction of drift force quantified by eigenvalues of the drift matrix $\Sigma$, by adding the coherence. Quantitatively it is $3-F|_{\epsilon=1}<3-F|_{\epsilon=0}$ in the region $2\Delta\ge \varepsilon_1-\varepsilon_2$, as estimated in the right figure in Fig.\ref{f2}. To elucidate this, we adopt the Langevin-Heisenberg theory \cite{Scully97} to develope the dynamical equation for the system operators $a_{\nu}$
\begin{equation}
\begin{split}
& \dot{a}_1 = -i\omega_1 a_1-iu a_2+\frac{1}{i\hbar}\sum_{\textbf{k},\sigma}g_{\textbf{k}\sigma}b_{\textbf{k}\sigma}^{(1)}\\
& \dot{a}_2 = -iu a_1-i\omega_2 a_2+\frac{1}{i\hbar}\sum_{\textbf{k},\sigma}\left(g_{\textbf{k}\sigma}b_{\textbf{k}\sigma}^{(1)}+f_{\textbf{k}\sigma}b_{\textbf{k}\sigma}^{(2)}\right)\\
& \dot{b}_{\textbf{k}\sigma}^{(1)} = -i\omega_{\textbf{k}\sigma}b_{\textbf{k}\sigma}^{(1)}+\frac{g_{\textbf{k}\sigma}}{i\hbar}\left(a_1+a_2\right)\\
& \dot{b}_{\textbf{k}\sigma}^{(2)} = -i\omega_{\textbf{k}\sigma}b_{\textbf{k}\sigma}^{(2)}+\frac{f_{\textbf{k}\sigma}}{i\hbar}a_2
\end{split}
\label{21}
\end{equation}
Eliminate the bath freedom by formally solving the last two equations in Eq.(\ref{21}) above
\begin{equation}
\begin{split}
& \begin{pmatrix}
   \dot{a}_1\\[0.2cm]
   \dot{a}_2\\
  \end{pmatrix}
= -\begin{pmatrix}
    i\omega_1+\gamma & iu+\gamma\\[0.2cm]
    iu+\gamma & i\omega_2+2\gamma\\
   \end{pmatrix}
   \begin{pmatrix}
    a_1\\[0.2cm]
    a_2\\
   \end{pmatrix}
  +\begin{pmatrix}
    F_1(t)\\[0.2cm]
    F_2(t)\\
   \end{pmatrix}\\[0.1cm]
& F_1(t) = \frac{1}{i\hbar}\sum_{\textbf{k},\sigma}g_{\textbf{k}\sigma}b_{\textbf{k}\sigma}^{(1)}(0)e^{-i\omega_{\textbf{k}\sigma}t}\\[0.1cm]
& F_2(t) = \frac{1}{i\hbar}\sum_{\textbf{k},\sigma}\left[g_{\textbf{k}\sigma}b_{\textbf{k}\sigma}^{(1)}(0)+f_{\textbf{k}\sigma}b_{\textbf{k}\sigma}^{(2)}(0)\right]e^{-i\omega_{\textbf{k}\sigma}t}
\end{split}
\label{22}
\end{equation}
which indeed is the quantum Langevin equation. $F_j(t)$ on the right hand side of Eq.(\ref{22}) stands for the stochastic force. As is known, the left hand side of Eq.(\ref{22}) represents the drift force in phase space and then the right hand side recovers the drift matrix defined before as quantifying the drift force induced by the environments. Therefore, {\it the dephasing originates from the drift force induced by the random scattering between the system and environmental modes. The effect of environment-induced coherence-population entanglment is to reduce the drift force and the coherence subsequently survives much longer}. On the other hand, it elucidates here that the environments have {\it non-trivial} contribution to the long-lived coherence in the excitation energy transport, contrary to the previous statements.

To explore the gradual effect of coherence-population entanglement on the drift force and the relaxation process, we need to further study the response of quantity $F$ to the adiabatic variation of the strength of the environment-induced coherence-population entanglement in QME. As shown in Fig.\ref{F}, the dephasing becomes rapid in the weak $\epsilon$ regime ($\epsilon\ll 1$) while it is reduced in the strong $\epsilon$ regime ($\epsilon \sim 1$), as the coherence-population entanglement increases. Physically this indicates that the coherence can reduce the environmental diffusion only when its coupling to population dynamics becomes significant. Such behavior can be understood in general by the following expansion of $F$ about $\epsilon=0$ and 1
\begin{equation}
\begin{split}
& F = F|_{\epsilon=0} + \left(1+\frac{1-(2d+w)^2}{\sqrt{(1-4d^2-w^2)^2+4w^2}}\right)\frac{\epsilon}{F|_{\epsilon=0}} + o(\epsilon^2)\\
& F = F|_{\epsilon=1} + \left(1+\frac{5+(6d+w)(2d-w)}{\sqrt{(5-4d^2-w^2)^2+4(4d-w)^2}}\right)\frac{\epsilon-1}{F|_{\epsilon=1}}\\
& \qquad\qquad\qquad\ + o\left[(\epsilon-1)^2\right]
\end{split}
\label{23}
\end{equation}
In our regime of parameters for vibrational energy transport, $\partial F/\partial\epsilon|_{\epsilon=0}$ is always negative while $\partial F/\partial\epsilon|_{\epsilon=1}$ is always positive.

Before leaving this section, it is worthwhile to point out that the mechanism of the increase of lifetime of coherence is not only restricted to the molecular vibrations as discussed here, but can also be applied to the exciton process in photosynthesis \cite{Palmieri09} described by the Redfield equation \cite{Breuer02}, where the coherence-population entanglement (beyond secular approximation) led to the enhancement of coherence lifetime as well. In this sense the reduction of decoherence caused by the environmental-induced coherence-population coupling is a general feature based on the structure of quantum master equation.

\section{Heat current}
Macroscopically, the energy transfer should be affected by the quantum interference, as being much debated in the excitation energy transport in light-harvesting complex. Here we will carefully explore the contribution by the non-local correlation originated from coherence, to the heat current. The transient process, or in other words, the relaxation, demands the non-vanishing energy consumption for the molecular mechine to reach the nonequilibrium steady state. The energy consumption in our model is therefore $Q=\int_0^{\infty}(\langle J_1\rangle-\langle J_2\rangle)dt = \int_0^{\infty}[\langle J_1\rangle-\langle J_1^{\infty}\rangle-(\langle J_2\rangle-\langle J_2^{\infty}\rangle)]dt$ and after some algebra the energy consumption reads
\begin{equation}
\begin{split}
& Q = Q_p-4\Delta\int_0^{\infty}\textup{Re}\left(\bar{C}[\rho]\right)dt\\
& \frac{Q_p}{2} = -\sum_{i=1}^2\left(\varepsilon_1\mathcal{I}_{ii}^{13}+\varepsilon_2\mathcal{I}_{ii}^{24}\right)\textup{Y}_i^i + \left(\varepsilon_1\mathcal{I}_{1221}^{13}+\varepsilon_2\mathcal{I}_{1221}^{24}\right)\textup{Y}_{12}^{21}\\
& \int_0^{\infty}\textup{Re}\left(\bar{C}[\rho]\right)dt = \textup{Re}[\mathcal{I}_{11}^{14}]\textup{Y}_1^1+\textup{Re}[\mathcal{I}_{22}^{14}]\textup{Y}_2^2+\textup{Re}[\mathcal{I}_{1221}^{14}]\textup{Y}_{12}^{21}
\end{split}
\label{25}
\end{equation}
where $\bar{C}[\rho]=C[\rho](t)-C[\rho](\infty)$. $\mathcal{I}_{...}^{...}=\gamma\int_0^{\infty}\mathcal{A}_{...}^{...}dt$ and their expressions will be given in SI. In general the 2$^{\textup{nd}}$ term in Eq.(\ref{25}) originates from the coherence, which demonstrates the non-negligible contribution from the coherence. To support this point numerically, Fig.\ref{F}(b) shows the {\it non-trivial} contribution of coherence-population entanglement to the energy consumption of the QHE. Moreover the coherence leads to the overall suppression of the energy consumption which in other words, indicates that the energy transport efficiency is effectively enhanced by adding the quantum interference.

\section{Conclusion and remarks}
We have studied the dynamical energy transport mediated by the molecular vibrations. It was found that the decohernece is much slower than the population relaxation, which suggests the coherent energy transfer, contrary to the one described by F$\ddot{\textup{o}}$rster theory. Since the quantum interference suppresses the drift force originated from the environments, the environment-induced coherence-population entanglement leads the coherence to survive much longer than the case with population involving only. Moreover the amplitudes of both coherence and population dynamics are also considerably amplified by the coherence-population entanglement. These demonstrate the significance of the {\it environment-assisted} coherence effect on the vibrational relaxation process. Our theoretical exploration further provides the prediction of the time scale of orientational relaxation of OH-stretching mode, with good agreement with the experimental measurements in HDO molecules dissolved in D$_2$O. On the macroscopic level, the coherence is shown to have {\it non-trivial} contribution to the enhancement of the quantum yield of vibrational energy transfer, as reflected by the suppression of the energy consumption.

\section{Acknowledgements}
We thank the support from the grant NSF-MCB-0947767.






\bibliography{rsc} 
\bibliographystyle{rsc} 

\end{document}